\documentclass[secnumarabic,amssymb, nobibnotes, aps, prl]{revtex4-2}

\usepackage{epsf}
\usepackage{amsfonts}
\usepackage{amssymb,yfonts,mathrsfs,bbm}
\usepackage{graphicx}
\makeatletter
\renewcommand{\fnum@figure}{Fig. \thefigure}
\makeatother
\usepackage{multirow}
\usepackage{longtable}
\usepackage{amsmath}
\usepackage[normalem]{ulem}
\usepackage{xcolor}
\usepackage{lineno}
\usepackage[colorlinks,
            linkcolor=red,
            anchorcolor=blue,
            citecolor=blue
            ]{hyperref}

\usepackage{siunitx}        % 物理单位规范排版
\usepackage{microtype}
\begin{document}
%\linenumbers

\title{Reconfigurable Chiral Phonons and Their Selection in Asymmetrically Stacked Graphene Moir\'{e} Superlattices}
%
%Raman tensor in thin in-plane anisotropic semiconductor flakes
%Quantitative prediction of Raman selection rule in thin in-plane anistropic black phoshporus/layered materials
\author{Yi-Fei Zhang$^{1,2,\ddag}$, Ting-Ting Wang$^{3,\ddag}$, Zhong-Jie Wang$^{1,2}$, Miao-Ling Lin$^{1,2}$, Yue Wang$^{1,2}$, Zi-Jian Zhong$^{1,2}$, Xue-Lu Liu$^{1,2}$, Heng Wu$^{1,2}$, Lin-Shang Chen$^{1,2}$, Yan-Meng Shi$^{1,2}$, Jin-Yang Lou$^{1,2}$, Lin-Han Li$^{1,2}$, Jia-Liang Xie$^{1,2}$, Rui Mei$^{1,2}$, Ti-Ying Zhu$^{1,2}$}
\author{Li-Fa Zhang$^{3}$}
\email{phyzlf@njnu.edu.cn}
\author{Jiang-Bin Wu$^{1,2}$}
\email{jbwu@semi.ac.cn}
\author{Ping-Heng Tan$^{1,2}$}
\email{phtan@semi.ac.cn}
\affiliation{$^1$ State Key Laboratory of Semiconductor Physics and Chip Technologies, Institute of Semiconductors, Chinese Academy of Sciences, Beijing, 100083, China}
\affiliation{$^2$ Center of Materials Science and Optoelectronics Engineering, University of Chinese Academy of Sciences, Beijing, 100049, China}
\affiliation{$^3$ Phonon Engineering Research Center of Jiangsu Province, Ministry of Education Key Laboratory of NSLSCS, Center for Quantum Transport and Thermal Energy Science, Institute of Physics Frontiers and Interdisciplinary Sciences, School of Physics and Technology, Nanjing Normal University, Nanjing, 210023, China}
\affiliation{$^\ddag$ Contributed equally to this work}
\maketitle

\section{Abstract}
Chiral phonons, lattice vibrations carrying angular momentum, offer a new handle for controlling phonon-mediated phenomena in condensed matter, yet their on-demand generation and deterministic control remain challenging. Here we show, using helicity-resolved Raman spectroscopy, that asymmetrically stacked graphene moir\'{e} superlattices provide a reconfigurable platform for chiral phonons. We observe a finite degree of circular polarization for both chiral interlayer shear phonons and intralayer optical phonons. A chiral-phonon Raman-tensor model attributes this response to interfacial $C_{2}$-symmetry breaking together with Raman resonance within the van Hove singularity region. Reversing either the twist sense or the stacking order inverts the degree of circular polarization through the spatial transformations linking the corresponding stacking geometries. By tuning the excitation energy and twist angle, we achieve nearly full-range control of the degree of circular polarization of interlayer shear phonons, from $-1$ to $1$, while gate-induced interlayer potential redistribution enables substantial and reversible in situ modulation. These results establish asymmetrically stacked graphene moir\'{e} superlattices as a versatile platform for generating and controlling chiral phonons, with implications for optoelectronic and quantum technologies.
\\

Keywords: chiral phonons, asymmetrically stacked graphene moir\'{e} superlattice, helicity-resolved Raman spectroscopy, van Hove singularities, reconfigurable phononics

\section{I. Introduction}
Chiral phonons arise from circular or elliptical lattice motion and possess both a real space angular momentum and a quantized pseudo-angular momentum (PAM)\cite{Zhang-PRL-2014, Zhang-PRL-2015, Zhu-Science-2018, Wang-NanoLett-2024, Juraschek-NaturePhysics-2025}. Recently, chiral phonons have been predicted in a wide range of lattice geometries, for example kagome\cite{Chen-PRB-2019}, Kekul\'{e}\cite{Liu-PRL-2017}, tetragonal\cite{Wang-PRB-2024}, and moir\'{e} lattices\cite{Xiao-NanoLett-2021}, and observed by various approaches, including transient infrared spectroscopy\cite{Zhu-Science-2018}, Raman spectroscopy\cite{Zhang-NanoLett-2023, Oishi-PRB-2024}, resonant inelastic X-ray scattering\cite{Ueda-Nature-2023}, and ultrafast electron diffraction\cite{Tauchert-Nature-2022}. They can also couple to other particles or quasiparticles such as electrons\cite{Tauchert-Nature-2022}, excitons\cite{Li-ACSNano-2019} and magnons\cite{Sasaki-NatCommun-2021, Choi-SciAdv-2022, Wu-NatPhys-2023}.

Since phonons govern charge, heat, and light transport in crystalline solids\cite{Huang-Book-1996}, chiral phonons were proposed to play key roles in many novel effects due to the unique chiral degree of freedom, such as high-temperature superconductivity\cite{Gao-PRB-2023}, directional thermal management\cite{Chen-NanoLett-2020}, and single-photon entanglement\cite{Chen-NatPhys-2019}. Realizing these proposals requires the on-demand generation and reversible control of chiral phonons. This is hindered in bulk materials by immutable lattice geometries that directly fix the properties of chiral phonons\cite{Chen-NanoLett-2020, Gao-PRB-2023}.
However, van der Waals heterostructures can overcome this limitation, because their dangling-bond-free interfaces eliminate lattice-mismatch constraints and enable precise control over lattice stacking geometry, thereby allowing tunability of chiral phonon properties.
In graphene moiré superlattices (GMSs), such highly tunable structures have already been shown to host van Hove singularities (vHSs)\cite{Li-NatPhys-2010, Brihuega-PRL-2012, Wu-NatCommun-2014}, correlated insulators\cite{Cao-Nature-2018, Codecido-SciAdv-2019}, quantum anomalous Hall states\cite{Serlin-Science-2020, Li-Nature-2021}, and unconventional superconductivity\cite{CaoYuan-Nature-2018, Codecido-SciAdv-2019, Yankowitz-Science-2019}.
Controlling the twist sense, twist angle and stacking order between two graphene sheets enables the construction of GMSs with programmable lattice chirality, in which circular superpositions of doubly degenerate in-plane modes can form oppositely rotating phonon states that couple selectively to circularly polarized light\cite{Chen-NatPhys-2019,Tatsumi-PRB-2018}. This structural tunability provides a direct handle to tailor chiral phonon properties, which remain largely unexplored.

Here we demonstrate, by helicity-resolved Raman spectroscopy (HRRS), that asymmetrically stacked GMSs constitute a reconfigurable platform for chiral phonons.
Asymmetric stacks exhibit a pronounced chirality-dependent Raman response for both chiral interlayer shear (C) and intralayer optical (G) phonons at the $\Gamma$ point of the moiré Brillouin zone, whereas symmetric stacks do not.
A chiral-phonon Raman-tensor analysis attributes this contrast to twist-induced \(C_{2}\) symmetry breaking together with Raman resonance within the vHS region.
Supported by first-principles calculations, this framework further explains why the degree of circular polarization (DoCP) is much larger for the C phonons than for the G phonons.
The spatial transformations relating the stacking geometries reveal that the DoCP can be reversibly controlled by tuning the interfacial twist sense and stacking order.
By tuning the excitation energy ($E_\text{ex}$), the interfacial twist angle ($\theta$), we achieve reversible control over both the sign and the magnitude of the DoCP for the C phonons, enabling nearly full-range tunability from $-1$ to $1$. In addition, electric field controlled interlayer potential redistribution enables substantial modulation of the DoCP.
These findings establish twist and stacking engineering as a versatile route for generating and controlling chiral phonons in van der Waals heterostructures and pave the way for their applications in optoelectronics, superconductivity, and quantum technologies.

\section{II. Results and Discussion}
\label{Results and discussion}
\subsection{Lattice Dynamics and Phonon Chirality in GMS}
\label{subsec: Lattice Dynamics and Phonon Chirality in GMS}

\begin{figure*}[htbp]
\centering
\includegraphics[width=0.7\textwidth]{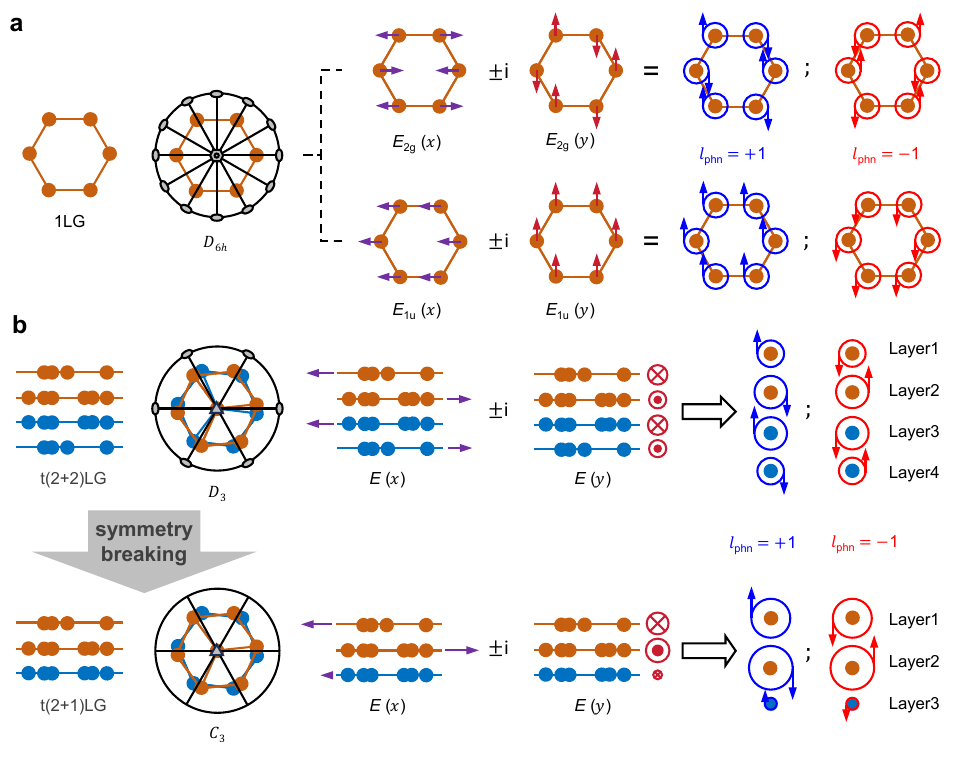}
\caption{Schematic diagrams of lattice structures and dynamics in 1LG and GMSs with basal planes in the x-y plane. (a) Lattice structure, stereographic projection, linear atomic displacements of the $E_\mathrm{1u}$ and $E_\mathrm{2g}$ modes in 1LG, and the resulting circular atomic motions constituting chiral phonons. (b) Lattice structure, stereographic projection, linear interlayer displacements of the $E$ modes, and the corresponding circular layer motions (represented by circular atomic trajectories within each layer) for the chiral C phonons in t(2+1)LG and t(2+2)LG. The arrow lengths indicate the displacement or motion amplitudes, and the PAM of the chiral phonons is indicated.}
\label{Schematic diagrams of lattice vibrations in 1LG and GMSs.}
\end{figure*}

Monolayer graphene (1LG) exhibits a planar hexagonal lattice with $D_\mathrm{6h}$ point-group symmetry\cite{Bernal-PRSA-1924, Novoselov-Science-2004} (Fig. 1a, left panel), where at the $\Gamma$ point of the  Brillouin zone, the doubly degenerate in-plane acoustic phonons correspond to the $E_\mathrm{1u}$ irreducible representation and the G phonons to $E_\mathrm{2g}$\cite{Tan-NatMater-2012}. In photon–phonon interactions dictated by angular momentum conservation, the degenerate orthogonal linear vibrations superpose to form symmetry-adapted circularly polarized chiral phonon modes that carry nonzero angular momentum (Fig. 1a, right panel)\cite{Zhang-PRL-2015, Chen-NatPhys-2019, Yin-AdvMater-2021, Minakova-NatPhys-2026}.

GMSs are heterostructures formed by twisting two constituents—either monolayer or AB-stacked multilayer graphene—by a specific angle (Fig. 1b, left panel).
Structurally, GMSs are categorized into symmetric stacks (e.g., t($n$+$n$)LG, consisting of two $n$-layer blocks) with $D_{3}$ point-group symmetry, and asymmetric stacks (e.g., t($m$+$n$)LG, with $m \neq n$) exhibiting $C_{3}$ symmetry\cite{Wu-ACSNano-2015}.
The lattice vibrations in GMSs inherit and extend those of monolayer graphene: at the $\Gamma$ point of the moiré Brillouin zone, besides retaining G and acoustic phonons, GMSs support finite-energy C phonons—derived from the acoustic phonons of 1LG—characterized by rigid-layer in-plane relative displacements\cite{Wu-NatCommun-2014}. Specifically, $n$LG hosts $n$-1 doubly degenerate C phonons, denoted C$_{n,n-j}$ ($j = 1, 2, \dots, n-1$), among which C$_{n,1}$ phonons have the highest frequency. Both C and G phonons in GMSs also exhibit chiral character under photon–phonon interactions (Fig. 1b, right panel). Due to the absence of $C_{2}$ symmetry at the twisted interface in t($m$+$n$)LG (unlike t($n$+$n$)LG), the vibrational amplitudes of C phonons are no longer correlated between the $m$LG and $n$LG constituents.
As illustrated in Fig. 1b, the layer displacements of the C$_{2,1}$ phonons in t(2+1)LG and t(2+2)LG, computed via the linear chain model\cite{Wu-NatCommun-2014}, show that C$_{2,1}$ phonons are localized within the 2LG components. The displacement amplitudes for linear and resulting circular motions in t(2+1)LG notably deviate from the mirror-symmetric ones observed in t(2+2)LG, consistent with the symmetry-breaking effect.

\subsection{Chirality-Dependent Raman Response in GMS}
\label{subsec: Chirality-Dependent Raman Response in GMS}

\begin{figure*}[htbp]
\centering
\includegraphics[width=0.80\textwidth]{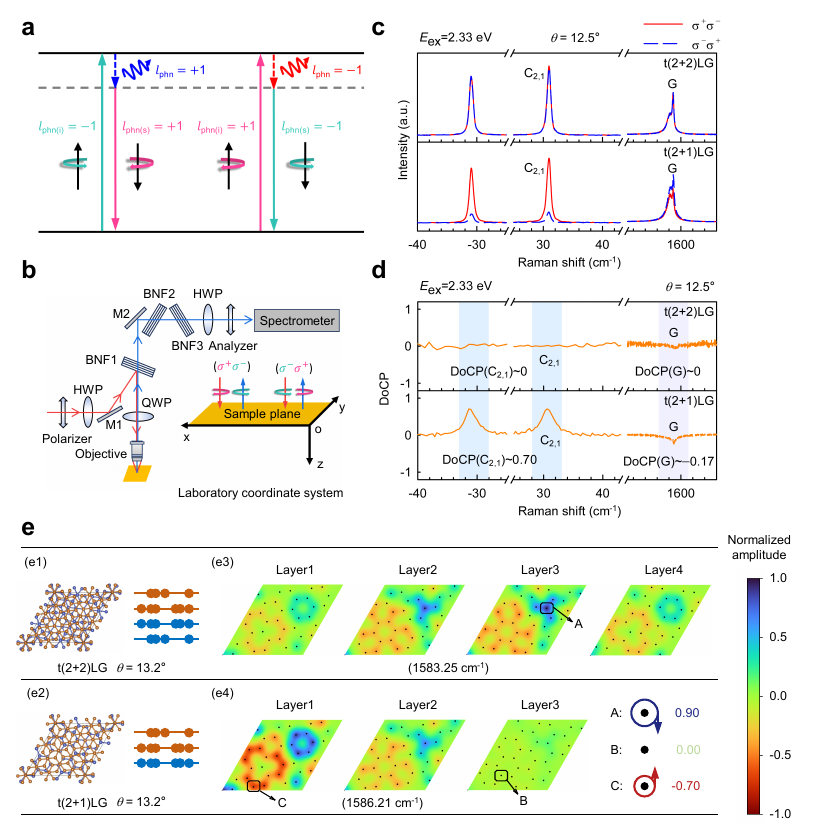}
\caption{HRRS of chiral phonons in GMSs.
(a) Schematic of the PAM conservation rule in the Raman process.
(b) Schematic diagram of the experimental setup of HRRS and circular polarization configurations in the laboratory coordinate system. Abbreviations: M, mirror; HWP, half-wave plate; QWP, quarter-wave plate; BNF, BragGrate notch filter.
(c) HRRS of the chiral C and G  phonons in t(2+2)LG and t(2+1)LG under $\sigma^{+}\sigma^{-}$ and $\sigma^{-}\sigma^{+}$ configurations excited by excitation energy of 2.33 eV.
(d) DoCP spectra of the chiral C and G  phonons in t(2+2)LG and t(2+1)LG with the corresponding DoCP values are labeled.
(e) First-principles, layer-resolved amplitude maps of circular atomic motions for the G phonons in (e3) t(2+2)LG ($\theta = 13.2^\circ$) and (e4) t(2+1)LG ($\theta = 13.2^\circ$). The lattice structures (top and side views) of t(2+2)LG and t(2+1)LG are shown in (e1) and (e2), respectively. The sign of the color indicates the handedness of the circular atomic motions, whereas the color intensity quantifies their relative amplitude. All color intensities are normalized to the maximum circular-motion amplitude among all G phonons observed in t(2+1)LG-CW (see Supplementary Materials Section I.3 for details). Representative points A, B, and C are highlighted together with the corresponding circular atomic motions and their quantitative values.
}
\label{HRRS of GMSs}
\end{figure*}

In GMSs, the $\Gamma$-point chiral phonons host intrinsic angular momentum confined to the basal plane, which is designated as the $x$-$y$ plane in our laboratory Cartesian coordinate system. Owing to the threefold rotational symmetry, these chiral phonons possess a quantized PAM ($l_{\text{phn}}$) of $\pm 1$, with the sign set by the sense of rotation viewed along the $z$-axis from the origin toward $+z$ (clockwise for $+1$, counterclockwise for $-1$; Supplementary Materials Note 1 for details). Photons in HRRS carry spin angular momentum ($l_{\text{pht}}$) of $\pm 1$, corresponding to an electric field vector rotating clockwise ($+1$) or counterclockwise ($-1$) within the same $x$-$y$ plane under the same viewing convention \cite{Oishi-PRB-2024}. PAM conservation during Raman scattering thus gives rise to the selection rules governing photons and $\Gamma$-point chiral phonons: $l_{\text{pht(s)}}-l_{\text{pht(i)}}=-l_{\text{phn}}+3N$, where $l_{\text{pht(i)}}$ and $l_{\text{pht(s)}}$ denote the $l_{\text{pht}}$ of incident and scattered photons, respectively, and $N$ is an integer arising from the threefold rotational symmetry of the lattice \cite{Zhang-PRL-2015, Tatsumi-PRB-2018, Yin-AdvMater-2021, Oishi-PRB-2024}. Consequently, HRRS with $l_{\text{pht(i)}}=-1$ and $l_{\text{pht(s)}}=+1$ selectively probes chiral phonons with $l_{\text{phn}}=+1$, while $l_{\text{pht(i)}}=+1$ and $l_{\text{pht(s)}}=-1$ select $l_{\text{phn}}=-1$ (Fig. 2a).

In HRRS experiments (see Fig. 2b and Methods for details), the circular polarization states of the incident and scattered light corresponding to photons with $l_{\text{pht}}=\pm 1$ are denoted as $\sigma^{+}$ and $\sigma^{-}$, respectively. This definition intentionally differs from the conventional left-/right-handed classification tied to the light-propagation direction (see Methods for details). The four HRRS polarization configurations ($\sigma^{+}\sigma^{+}$, $\sigma^{-}\sigma^{-}$, $\sigma^{+}\sigma^{-}$, and $\sigma^{-}\sigma^{+}$) are defined by listing the helicities of the incident and scattered photons in order. As shown in Supplementary Fig. S1, the G phonons in 1LG exhibit Raman signals exclusively under the $\sigma^{+}\sigma^{-}$ and $\sigma^{-}\sigma^{+}$ configurations. This observation is fully consistent with PAM conservation and prior reports\cite{Drapcho-PRB-2017}, confirming the validity of our experimental setup for probing phonon chirality.

The HRRS spectra for t(2+2)LG and t(2+1)LG ($\theta = 12.5^{\circ}$) are presented in Fig. 2c. The prominent peaks at $\sim$ 31 cm$^{-1}$ and $\sim$ 1580 cm$^{-1}$ in both samples correspond to the C$_{2,1}$  and G phonons, respectively, as identified by phonon dispersion from first-principles calculations (Supplementary Fig. S2) and prior studies\cite{Wu-NatCommun-2014}. These phonons are observed under cross-circular polarization ($\sigma^{+}\sigma^{-}$ and $\sigma^{-}\sigma^{+}$) configurations, consistent with their character of chiral phonons. Notably, in t(2+2)LG the Raman intensity in the \(\sigma^{+}\sigma^{-}\) configuration \((I_{\sigma^{+}\sigma^{-}})\) equals that in the \(\sigma^{-}\sigma^{+}\) configuration \((I_{\sigma^{-}\sigma^{+}})\), whereas in t(2+1)LG the two configurations yield different intensities for both the C and G phonons. This intensity asymmetry is a clear signature of chirality-dependent Raman response, which can be quantified as DoCP:

\begin{equation}
\label{eq:DOCP_def}
\mathrm{DoCP}
= \frac{I_{\sigma^{+}\sigma^{-}} - I_{\sigma^{-}\sigma^{+}}}{I_{\sigma^{+}\sigma^{-}} + I_{\sigma^{-}\sigma^{+}}}.
\end{equation}

\noindent At $E_\text{ex}=2.33$ eV, the C$_{2,1}$ phonons exhibit a large DoCP of 0.70, whereas the G phonons show a much smaller value of $-0.17$ (Fig. 2d).

\subsection{Chiral-Phonon Raman-Tensor Model}
\label{subsec: Chiral phonon Raman Tensor Model}

To understand the origin of the chirality-dependent Raman response, we derive the relation between the Raman tensors of the chiral phonons and those of their orthogonal linear counterparts. As illustrated by the atomic motions in Fig. 1, the normal coordinates ($Q_{\pm}$) for phonons with $l_{\text{phn}}=\pm 1$ can be written as ${Q}_{\pm}=(Q_1 \pm \text{i}Q_2)/\sqrt{2}$, where $Q_1$ and $Q_2$ are the normal coordinates of the corresponding orthogonal linear vibrational modes. The Raman tensor $\mathbf{R}_{\nu}$ for a vibrational mode $\nu$ is defined as the derivative of the polarizability $\chi$ with respect to its normal coordinate, \mbox{$\partial \chi / \partial Q_{\nu}$} evaluated at $Q=0$\cite{Loudon-AdvPhys-1964}. Accordingly, if $\mathbf{R}_1$ and $\mathbf{R}_2$ are the Raman tensors of the two orthogonal linear modes, the Raman tensors for the chiral phonons with $l_{\text{phn}}=\pm 1$ are given by $\mathbf{R}_{\pm}=(\mathbf{R}_{1} \mp \text{i}\mathbf{R}_{2})/\sqrt{2}$ (see Supplementary Materials Note 2 for details). The Raman intensities of the chiral phonons under circular-polarization configurations are therefore $I_{\pm} \propto |\mathbf{e}_s^{\dagger} \cdot \mathbf{R}_{\pm} \cdot \mathbf{e}_i|^2$, where $\mathbf{e}_i$ ($\mathbf{e}_s$) denotes the Jones vector of the incident (scattered) light with circular polarization state $\sigma^{+}$ or $\sigma^{-}$, explicitly written as $\frac{1}{\sqrt{2}}(1, \text{i}, 0)^{\text{T}}$ or $\frac{1}{\sqrt{2}}(1, -\text{i}, 0)^{\text{T}}$, respectively, and $\mathbf{e}_s^{\dagger}$ is the Hermitian conjugate of $\mathbf{e}_s$. This chiral-phonon Raman-tensor model reproduces the Raman intensities observed experimentally while preserving PAM conservation. Under the $\sigma^{+}\sigma^{-}$ and $\sigma^{-}\sigma^{+}$ configurations, the G phonons with $l_{\text{phn}}=\mp 1$ in 1LG, whose orthogonal linear counterparts belong to the $E_\mathrm{2g}$ irreducible representation, are selectively excited. The Raman tensors $\mathbf{R}_{-}(E_\mathrm{2g})$ and $\mathbf{R}_{+}(E_\mathrm{2g})$ are Raman active only under the $\sigma^{+}\sigma^{-}$ and $\sigma^{-}\sigma^{+}$ configurations, respectively, and yield equal Raman intensities. This leads to the intensity ratio $I_{\sigma^{+}\sigma^{-}}:I_{\sigma^{-}\sigma^{+}}:I_{\sigma^{+}\sigma^{+}}:I_{\sigma^{-}\sigma^{-}}=1:1:0:0$ (see Supplementary Note 3 for details). These results reproduce the experimental HRRS behavior shown in Supplementary Fig. S1 and validate the proposed theoretical framework for chiral phonons.

All C and G phonons in both t($n$+$n$)LG and t($m$+$n$)LG belong to the $E$ irreducible representation. However, $C_2$ symmetry breaking at the twist interface reduces the point group symmetry from $D_3$ in symmetric t($n$+$n$)LG to $C_3$ in asymmetric t($m$+$n$)LG, which introduces additional nonzero elements in the Raman tensors of the $E$ phonons in t($m$+$n$)LG compared to those in t($n$+$n$)LG, as detailed below\cite{Loudon-AdvPhys-1964}:

\begin{equation}
\begin{aligned}
\mathbf{R}_{1,2}(E)~\text{for}~D_3:
\begin{bmatrix}
c & 0 & 0 \\
0 & -c & d \\
0 & d & 0
\end{bmatrix},
\begin{bmatrix}
0 & -c & -d \\
-c & 0 & 0 \\
-d & 0 & 0
\end{bmatrix}
,
\mathbf{R}_{1,2}(E)~\text{for}~C_3:
\begin{bmatrix}
c & e & f \\
e & -c & d \\
f & d & 0
\end{bmatrix},
\begin{bmatrix}
e & -c & -d \\
-c & -e & f \\
-d & f & 0
\end{bmatrix}.
\label{the Raman tensors of E}
\end{aligned}
\end{equation}

\noindent where all elements ($c$, $d$, $e$, and $f$) may be complex under resonant excitation \cite{Pimenta-PCCP-2021, Han-JPCL-2022, Mei-Nanoscale-2024}.
One can derive that, for t(2+2)LG,  $I_{\sigma^{+}\sigma^{-}}[\mathbf{R}_{-}(E)] = I_{\sigma^{-}\sigma^{+}}[\mathbf{R}_{+}(E)] \propto 2|c|^2$ and $\mathrm{DoCP}=0$. In contrast, for t(2+1)LG, the Raman intensities yield different expressions for the $\sigma^{+}\sigma^{-}$ and $\sigma^{-}\sigma^{+}$ configurations, i.e., $I_{\sigma^{+}\sigma^{-}}[\mathbf{R}_{-}(E)] \propto 2|c+e\text{i}|^2$ and  $I_{\sigma^{-}\sigma^{+}}[\mathbf{R}_{+}(E)] \propto 2|c-e\text{i}|^2$, and the corresponding DoCP of the C and G phonons takes the form:

\begin{equation}
\label{eq:DOCP_ce}
\mathrm{DoCP}
=\frac{2|c+e\text{i}|^2 - 2|c-e\text{i}|^2}{2|c+e\text{i}|^2 + 2|c-e\text{i}|^2}
= \frac{2\,\mathrm{Im}(c\,e^{*})}{\lvert c\rvert^{2}+\lvert e\rvert^{2}}.
\end{equation}

\noindent Obviously, the observation of a nonzero DoCP in GMSs necessitates nonzero $c$ and $e$ with a relative phase difference. Unlike t(2+2)LG, the interface $C_2$ symmetry breaking in asymmetrically stacked GMS introduces a nonzero $e$. Previous studies on GMSs at $\theta = 12.5^\circ$ have indicated that the C$_{2,1}$ phonons in t(2+1)LG are resonantly excited within the vHS range at $E_\text{ex} = 2.33$ eV \cite{Carozo-PRB-2013}, introducing complex values for the Raman tensor elements $c$ and $e$. Consequently, a nonzero DoCP is observed for the C$_{2,1}$ phonons in t(2+1)LG, in contrast to the zero DoCP observed in t(2+2)LG, as demonstrated in Fig. 2c and Fig. 2d.

\begin{figure*}[htp!]
\centering
\includegraphics[width=1\textwidth]{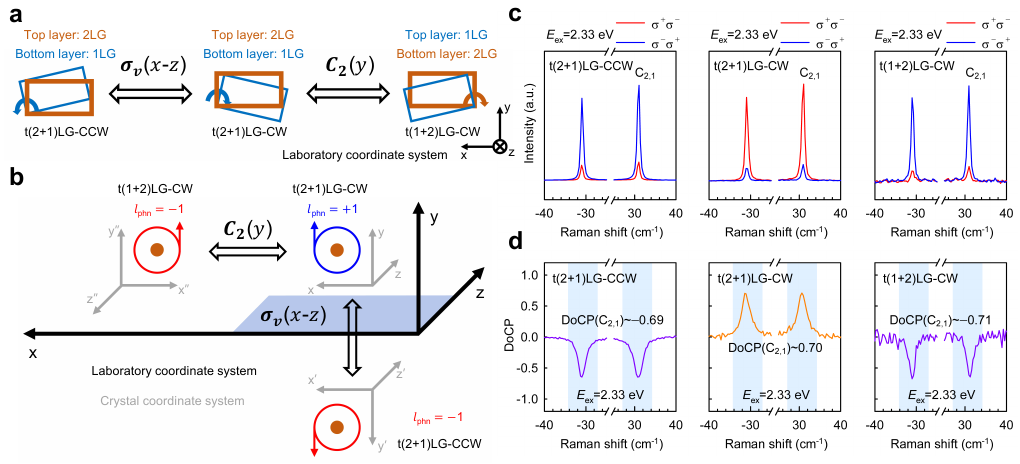}
\caption{ Relationship between DoCP and stacking geometries in asymmetrically stacked GMSs. (a) Schematic diagram of stacking configuration and spatial transformation relationships among t(2+1)LG-CCW, t(2+1)LG-CW, and t(1+2)LG-CW with identical twist angle. $\sigma_v$($x$-$z$) is the mirror inversion where the $x$-$z$ plane serves as the mirror plane and $C_{2} (y)$ is the twofold rotation about the $y$ axis, respectively.
(b) Schematic diagram of the relationships between coordinate systems and the phonon chirality. The laboratory coordinate system $(x,y,z)$ is depicted in black, while the crystal coordinate system $(x, y, z)$ of t(2+1)LG-CW, aligned with the laboratory frame, is shown in gray. The other two gray coordinate systems, $(x^{\prime}, y^{\prime}, z^{\prime})$ and $(x^{\prime\prime}, y^{\prime\prime}, z^{\prime\prime})$, are the crystal coordinate systems of t(2+1)LG-CW under the spatial transformations $\sigma_v$ ($x$-$z$) and $C_{2} (y)$, respectively. The PAM of the chiral phonons in the laboratory coordinate system is indicated.
(c) HRRS of the C$_{2,1}$ phonons in t(2+1)LG-CCW, t(2+1)LG-CW, and t(1+2)LG-CW (all with $\theta = 12.5^\circ$) at $E_{\text{ex}} = 2.33$ eV under $\sigma^{+}\sigma^{-}$ and $\sigma^{-}\sigma^{+}$ configurations.
(d) DoCP spectra of the C$_{2,1}$ phonons among different stacking geometries with the corresponding DoCP values are labeled.
}
\label{The CDRR of C phonons in t(m+n)LG}
\end{figure*}

To elucidate why the DoCP of the G phonons is much smaller than that of the C$_{2,1}$ phonons in t(2+1)LG, we calculated the two orthogonal linear vibration patterns of the G phonons in both t(2+2)LG and t(2+1)LG (Fig. 2(e1) and Fig. 2(e2)) using first-principles calculations, and then constructed their chiral superpositions (see Supplementary Materials Section I.3 for details). Fig. 2(e3) and Fig. 2(e4) visualize the normalized amplitudes of circular atomic motions associated with the G phonons in t(2+2)LG and t(2+1)LG, respectively. The comparison results reveal that breaking the $C_2$ symmetry fundamentally redistributes the phonon amplitude between layers but nearly preserves the intralayer motion pattern. The chiral G phonons in t(2+1)LG retain almost the same atomic vibrations in the plane as in t(2+2)LG by comparing the vibration amplitude distribution within the moir\'{e} superlattice (see Supplementary Fig. S3 and Fig. S4). Since the G phonons are predominantly governed by intralayer motions while the C phonons are strongly influenced by interlayer coupling\cite{Tan-NatMater-2012, Wu-NatCommun-2014}, the $C_2$ symmetry breaking introduces only very small changes to the Raman tensors of G phonons in t(2+1)LG, and thus $|e| \ll |c|$. This accounts for the much smaller DoCP observed for the G phonons than for the C$_{2,1}$ phonons in t(2+1)LG according to Eq.~\ref{eq:DOCP_ce}.

\subsection{Stacking Geometry Dependence of DoCP Sign}
\label{subsec: Stacking Geometry}

For fixed twist angle and layer number, an asymmetrically stacked GMS with basal planes in the $x$-$y$ plane exhibits four inequivalent stacking geometries connected by a twofold rotation $C_{2}$ about the $y$ axis (denoted as $C_{2}(y)$ in Fig. 3a), a mirror operation $\sigma_v$ across the $x$-$z$ plane (denoted as $\sigma_{v}(x\text{-}z)$ in Fig. 3a), and their combinations. As illustrated in Fig. 3a for the case $m=2$, $n=1$, and $\theta = 12.5^\circ$, t(2+1)LG-CW, namely t(2+1)LG with a clockwise twist of the bottom 1LG relative to the top 2LG, is transformed into its counterclockwise enantiomer t(2+1)LG-CCW by $\sigma_{v}(x\text{-}z)$. Applying $C_{2}(y)$ to t(2+1)LG-CW exchanges the 2LG and 1LG constituents and produces t(1+2)LG-CW. Applying the combined operation $\sigma_{v}(x\text{-}z)C_{2}(y)$ or $C_{2}(y)\sigma_{v}(x\text{-}z)$ to t(2+1)LG-CW yields t(1+2)LG-CCW. The transformation matrices of the $\sigma_v(x\text{-}z)$ and $C_{2}(y)$ operations shown in Fig. 3a.

\begin{equation}
\mathbf{M}_{\sigma_v (x\text{-}z)}
=
\begin{bmatrix}
1 & 0 & 0 \\
0 & -1 & 0 \\
0 & 0 & 1
\end{bmatrix},
\label{sigma-v x-z matrix}
\mathbf{M}_{C_2 (y)}
=
\begin{bmatrix}
-1 & 0 & 0 \\
0 & 1 & 0 \\
0 & 0 & -1
\end{bmatrix}.
\end{equation}

We performed HRRS on samples with the three representative configurations shown in Fig. 3a. As shown in Fig. 3c,d, the C$_{2,1}$ phonons in t(2+1)LG-CCW and t(2+1)LG-CW at the same twist angle of $12.5^\circ$ exhibit almost identical DoCP magnitudes but opposite signs. A similar sign reversal is observed between t(2+1)LG-CW and t(1+2)LG-CW. This behavior can be understood from the action of the $\sigma_v(x$-$z)$ and $C_{2}(y)$ operations on the phonon chirality and Raman tensors. As shown in Fig. 3b, under $\sigma_v(x$-$z)$, PAM reverses sign in the transformed coordinate system; for example, $l_{\text{phn}}=+1$ in t(2+1)LG-CW becomes $l_{\text{phn}}=-1$ in t(2+1)LG-CCW. Correspondingly, the Raman tensors of phonons with $l_{\text{phn}}=\pm 1$ in the CW and CCW stacks are related by $\mathbf{R}_{\mathrm{CCW}(\mp)} =
\mathbf{M}_{\sigma_v(x\text{-}z)}\,
\mathbf{R}_{\mathrm{CW}(\pm)}\,
\mathbf{M}_{\sigma_v(x\text{-}z)}^{\text{T}}$.
This relation leads directly to the Raman-intensity ratio $I_{\sigma^{+}\sigma^{-}}^{\mathrm{CW}} :
I_{\sigma^{-}\sigma^{+}}^{\mathrm{CW}} =
I_{\sigma^{-}\sigma^{+}}^{\mathrm{CCW}} :
I_{\sigma^{+}\sigma^{-}}^{\mathrm{CCW}}$, and hence to a sign reversal of the DoCP between CW and CCW stacks throughout the GMS family. The observed inversion of the DoCP in t($n$+1)LG-CW and t($n$+1)LG-CCW ($n = 3$ and $4$; Supplementary Fig. S5) further confirms this relation. Likewise, the $C_{2}(y)$ operation converts t($m$+$n$)LG into t($n$+$m$)LG while preserving the clockwise sense, leading to a similar DoCP reversal, consistent with the spectra of t(2+1)LG-CW and t(1+2)LG-CW in Fig. 3c,d. Because the DoCP is odd under both $\sigma_v(x$-$z)$ and $C_{2}(y)$, applying the combined operation $\sigma_{v}(x\text{-}z)C_{2}(y)$ or $C_{2}(y)\sigma_{v}(x\text{-}z)$ induces two sign reversals and therefore leaves the DoCP unchanged. As a result, t($m$+$n$)LG-CW and t($n$+$m$)LG-CCW with the same twist angle exhibit identical DoCP. These findings establish that the DoCP sign in asymmetrically stacked GMSs is governed by the spatial transformations relating their stacking geometries.

\begin{figure*}[htbp]
\centering
\includegraphics[width=0.96\textwidth]{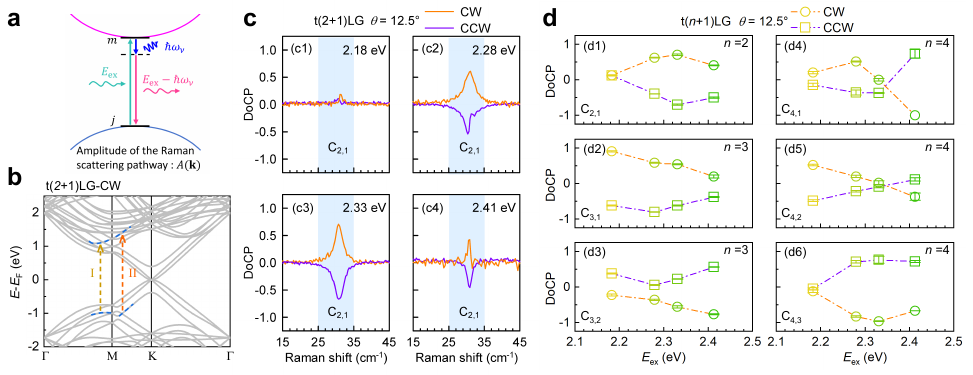}
\caption{Excitation energy dependence of the DoCP in asymmetrically stacked GMSs. (a) Schematic illustration of the Raman scattering pathway of the first-order Raman process, $j$ and $m$ represent the electron states in the band structure. $\hbar\omega_{\nu}$ corresponding to the phonon energy of the phonon mode $\nu$. (b) Calculated band structure of t(2+1)LG-CW ($\theta = 13.2^{\circ}$). The arrows (I and II) represent distinct optical transition pathways within the vHS energy range in the joint density of states.
(c) DoCP spectra of the C$_{2,1}$ phonons in t(2+1)LG-CW and t(2+1)LG-CCW with excitation energy of (c1) 2.18 eV, (c2) 2.28 eV, (c3) 2.33 eV, and (c4) 2.41 eV.
(d) Excitation energy dependence of the DoCP for C phonons in the t($n+1$)LG-CW and t($n+1$)LG-CCW under different excitation energy, $n$ = 2, 3, and 4. The colored circular markers connected by dashed lines (guide lines) correspond to the experimental measurements, where the color of each marker encodes the excitation energy (2.18 eV, 2.28 eV, 2.33 eV, and 2.41 eV). Error bars represent the propagated combined uncertainty of DoCP values due to sample inhomogeneity and measurement error, based on two independent sites per sample.
}
\label{Dispersion characteristics of DoCP of the C phonons}
\end{figure*}

\subsection{DoCP Modulation by Excitation Energy}
\label{subsec: Excitation Energy}

The above analysis shows that the substantial DoCP observed in t($m$+$n$)LG originates from complex Raman tensor elements generated under resonant excitation within the vHS region. Within third-order perturbation theory, the intensity of a first-order Raman process is given by the squared modulus of the coherent sum of all scattering amplitudes over $k$ space, denoted as $A(\mathbf{k})$ in Fig.~4a.\cite{Loudon-AdvPhys-1964, Chen-Nature-2011, Chen-ACSNano-2023, Hung-CPC-2024} In t($m$+$n$)LGs, this sum is dominated by pathways involving optical transitions within the vHS region in the joint density of states,\cite{Wu-NatCommun-2014, Wu-ACSNano-2015} such as pathways I and II in Fig.~4b. As a result, even a small change in excitation energy across the resonance range can substantially reshape the coherent Raman response, leading to pronounced changes in the relative magnitudes and phases of the complex Raman tensor elements (see Supplementary Materials Note 4 for details). According to Eq.~\ref{eq:DOCP_ce}, these variations directly determine the DoCP of the chiral phonons, especially for the C phonons in asymmetrically stacked GMSs.

To test this picture, we performed HRRS measurements on t($m$+$n$)LGs at four excitation energies, 2.18, 2.28, 2.33 and 2.41~eV, spanning the vHS resonance region. As shown in Supplementary Fig.~S6, the Raman intensities of all C phonons vary strongly with excitation energy. For example, in t(2+1)LG-CW ($\theta = 12.5^\circ$), the $\sigma^{+}\sigma^{-}$ intensity of the C$_{2,1}$ phonon at 2.28~eV is about 30 times larger than that at 2.41~eV. The DoCP is likewise highly sensitive to excitation energy. In the same sample, the DoCP of the C$_{2,1}$ phonons is nearly zero at 2.18~eV but increases to 0.70 at 2.33~eV (Fig.~4c). In t(4+1)LG-CW ($\theta = 12.5^\circ$), the DoCP of the C$_{4,1}$ and C$_{4,2}$ phonons reverses sign, whereas that of the C$_{4,3}$ phonon evolves from nearly zero to nearly unity across 2.18--2.41~eV, consistent with the richer resonant electronic structure of this thicker stack (Fig.~S7). Moreover, at each excitation energy, samples with opposite twist senses exhibit nearly opposite DoCP values, as expected from the symmetry relation between the corresponding stacking geometries established above. These results show that tuning excitation energy within the vHS resonance range effectively modulates both the Raman intensity and the DoCP of chiral phonons, providing a direct optical handle for continuously reconfiguring chiral-phonon responses in asymmetrically stacked GMSs.

\subsection{DoCP Modulation by Band Engineering}
\label{subsec: Band Engineering}

\begin{figure*}[htbp]
\centering
\includegraphics[width=0.96\textwidth]{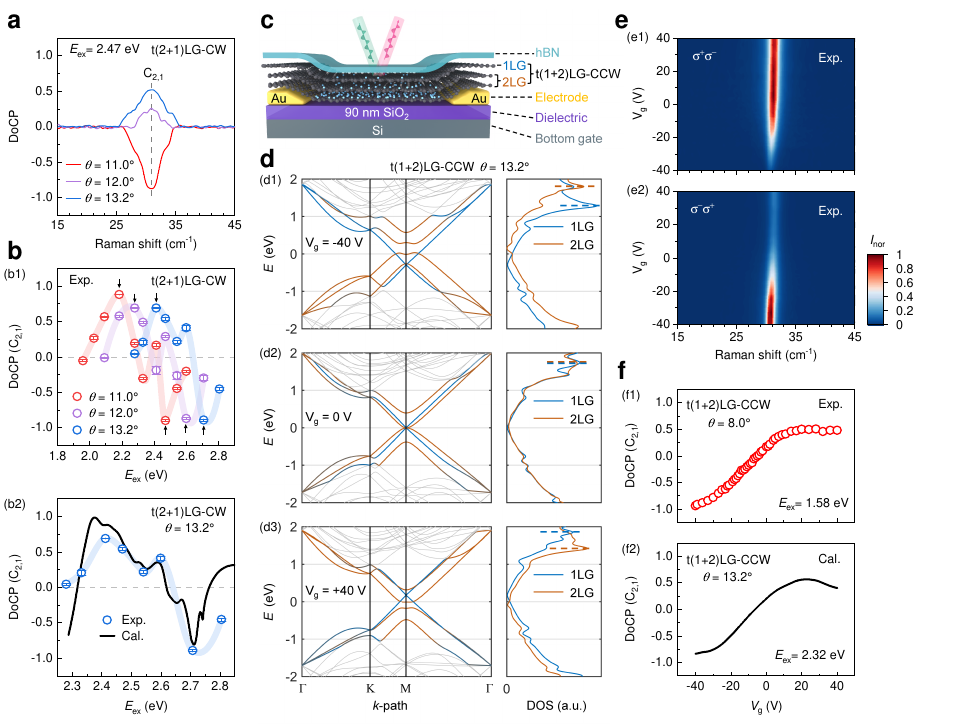}
\caption{Band engineering for DoCP modulation in t($m+n$)LGs.
(a) DoCP spectra of the C$_{2,1}$ phonon in t(2+1)LG-CW at the $\theta$ of $11.0^{\circ}$, $12.0^{\circ}$, and $13.2^{\circ}$, respectively.
(b) DoCP profile as a function of excitation energy within the vHS resonance region. (b1) DoCP profile of the C$_{2,1}$ phonon in t(2+1)LG-CW at $\theta = 11.0^{\circ}$ (red circles), $12.0^{\circ}$ (purple circles), and $13.2^{\circ}$ (blue circles). Error bars represent the propagated combined uncertainty of the DoCP due to sample inhomogeneity and measurement error, based on three independent sites per sample. Semi-transparent guide lines illustrate the evolution of the DoCP with excitation energy. Arrows mark the positions of the maximum and minimum DoCP values for each twist angle. (b2) Comparison between the experimental data (Exp., blue circles) and the calculated results (Cal., solid black line) for t(2+1)LG-CW at $\theta = 13.2^{\circ}$.
(c) Schematic of a field-effect transistor for electric-field tuning of the band structure of t($m+n$)LGs. Taking t(1+2)LG-CCW as an example, the figure illustrates the layer-resolved carrier distribution under a positive $V_{\text{g}}$.
(d) Band structures and density of states (DOS) of t(1+2)LG-CCW ($\theta = 13.2^\circ$) under applied $V_{\text{g}}$ of (d1) $-40$ V, (d2) $0$ V, and (d3) $+40$ V.
(e) Gate voltage dependent Raman intensity mapping (experimental) for t(1+2)LG-CCW ($\theta = 8.0^\circ$)  under (e1) $\sigma^{+}\sigma^{-}$ and (e2) $\sigma^{-}\sigma^{+}$ configurations.
The color bar on the right represents the normalized Raman intensity ($I_{\text{nor}}$), with red/blue indicating high/low intensity values, respectively.
(f) DoCP of the C\(_{2,1}\) phonon as a function of $V_{\text{g}}$ in t(1+2)LG-CCW. (f1) Experimental results for t(1+2)LG-CCW ($\theta = 8.0^\circ$) at $E_{\text{ex}} = 1.58$ eV; (f2) Calculation results for t(1+2)LG-CCW ($\theta = 13.2^\circ$) at $E_{\text{ex}} = 2.32$ eV.
}
\label{Band engineering}
\end{figure*}

In t($m+n$)LGs, the twist angle provides a powerful knob for tuning the resonant electronic structure. Its microscopic origin lies in the twist angle dependent moir\'{e} superlattice generated by twisting adjacent graphene constituents, which produces twist angle dependent vHS and optical resonance region.\cite{Li-NatPhys-2010, MacDonald-PNAS-2011, Brihuega-PRL-2012} Within the third-order perturbation Raman framework, these twist angle dependent vHSs reshape the resonant Raman pathways and thereby modify the magnitudes and relative phases of the Raman tensor elements. Eq.~\ref{eq:DOCP_ce} therefore implies that the DoCP of the C phonons should be tunable through twist angle. To test this expectation, we prepared t(2+1)LG-CW samples with $\theta = 11.0^\circ$, $12.0^\circ$, and $13.2^\circ$, and measured the corresponding DoCP at $E_{\text{ex}} = 2.47$ eV (Fig.~5a). As twist angle increases from $11.0^\circ$ to $13.2^\circ$, the DoCP of the C$_{2,1}$ phonons changes from $-0.90$ to 0.54. When excitation energy is further scanned across the vHS region, all three samples display a similar DoCP evolution, whereas the entire DoCP profile undergoes a nearly systematic blueshift with increasing twist angle (Fig.~5(b1)). This behavior is consistent with a shift of the underlying vHS resonance range and shows that the DoCP profile directly tracks the twist angle dependent resonant electronic structure. The same trend persists in thicker asymmetric stacks: the DoCP profiles of the C$_{4,1}$, C$_{4,2}$, and C$_{4,3}$ phonons in t(4+1)LG-CW ($\theta = 13.2^\circ$) are all blueshifted relative to those at $\theta = 12.5^\circ$ (Fig.~S7). Moreover, the calculated DoCP--$E_{\mathrm{ex}}$ profiles, obtained from the tight-binding calculations (see Methods), agree well with the experimental results for t(2+1)LG-CW and t(4+1)LG-CW at $\theta = 13.2^\circ$ (Fig.~5(b2) and Fig.~S8), supporting the picture that twist angle tunes the DoCP primarily by shifting the moir\'{e}-induced vHSs and the associated resonant Raman pathways.

The resonant electronic structure can also be tuned electrically. Fig.~5c shows a field-effect transistor based on t(1+2)LG-CCW. Because graphene screens a perpendicular electric field strongly but not completely, an applied gate voltage ($V_{\text{g}}$) induces unequal carrier densities in different graphene layers, thereby generating a layer-dependent interlayer potential distribution.\cite{Koshino-PRB-2014, Zhou-NC-2024} In the calculations, this effect is incorporated by adding a layer-dependent static potential, namely a Stark-effect term, to the zero-gate Wannier Hamiltonian (see Methods). This electric-field-induced potential redistribution shifts the relative alignment of the 1LG- and 2LG-derived bands and reconstructs the resonance structure. Fig.~5d presents the calculated band structures and corresponding density of states for t(1+2)LG-CCW ($\theta = 13.2^\circ$) at $V_{\text{g}} = -40$, 0, and $+40$ V. The band components associated with the 1LG and 2LG constituents exhibit a pronounced relative shift under gate bias, which leads to a splitting of the corresponding vHS features. This analysis indicates that gate voltage can effectively modulate both the Raman intensity and the DoCP of the C phonons.

This picture is supported by gate-dependent HRRS measurements on the t(1+2)LG-CCW field-effect transistor. At $E_{\mathrm{ex}} = 1.58$~eV, the Raman intensity of the C$_{2,1}$ phonon shows a pronounced gate voltage dependence under both $\sigma^{+}\sigma^{-}$ and $\sigma^{-}\sigma^{+}$ configurations, and the extracted DoCP evolves from $-0.9$ to 0.5 as gate voltage is swept from $-40$ to $+40$~V (Fig.~5e and Fig.~5(f1)). The accompanying small phonon-frequency shift is consistent with gate-modified interlayer coupling through carrier doping.\cite{Yuan-AMI-2020} For comparison, the gate-dependent DoCP was calculated for a representative commensurate t(1+2)LG-CCW structure with $\theta = 13.2^\circ$ at $E_{\mathrm{ex}} = 2.32$~eV (Fig.~5(f2)). Although the experimental and calculated cases differ in absolute twist angle and excitation energy, they were intentionally chosen to lie in closely analogous resonance regimes, as evidenced by their very similar zero-field DoCP values (see Fig.~S9). Under these conditions, the calculated DoCP of the C$_{2,1}$ phonon as a function of gate voltage agrees well with the experimental trend. Together, these results show that band engineering, through either twist-angle-controlled vHS resonance or electric-field-induced band reconstruction, provides an effective route for reconfiguring the response of chiral C phonons in asymmetrically stacked GMSs.

\section{III. Conclusion}
\label{Conclusion}

In summary, the chiral phonons in asymmetrically stacked GMSs are highly reconfigurable through stacking-geometry design, resonant-excitation selection, and band engineering. HRRS measurements on t($m{+}n$)LGs reveal a pronounced DoCP for the interlayer chiral C phonons, whereas t($n{+}n$)LGs show no such effect. A chiral-phonon Raman-tensor analysis, supported by first-principles calculations, identifies the synergistic roles of interfacial $C_2$-symmetry breaking and Raman resonance within the vHS region in generating the DoCP of C phonons, and explains why the interlayer C phonons respond much more strongly than the intralayer-dominated G phonons. By varying the excitation energy and twist angle, the DoCP can be tuned across nearly the full range from $-1$ to $1$. In addition, electric-field gating provides substantial and reversible in situ modulation through reconstruction of the resonant electronic structure. These results establish asymmetrically stacked GMSs as a reconfigurable platform for controlling chiral phonons and suggest new opportunities for chiral-phonon-based functionalities in optoelectronic and quantum devices.

\section{IV. Methods}
Additional experimental details including Raman measurements, sample and device preparation, and computational methodology are available in the Supplementary Materials.

\end{document}